\documentclass[aps,prb,reprint, superscriptaddress,secnumarabic,amssymb,nobibnotes]{revtex4-1}
\usepackage{graphicx}
\usepackage{amsmath}
\usepackage{subfigure}
\usepackage{verbatim}
\usepackage[utf8]{inputenc}
\usepackage{dsfont}
\usepackage[colorinlistoftodos]{todonotes}
\usepackage[version=3]{mhchem}
\usepackage{chemfig}
\usepackage{cancel}
\usepackage{hyperref}
\usepackage{xcolor}
\usepackage{upgreek}
\usepackage{bm}
\usepackage{amssymb}

\makeatletter
\newsavebox{\@brx}
\newcommand{\llangle}[1][]{\savebox{\@brx}{\(\m@th{#1\langle}\)}%
\mathopen{\copy\@brx\kern-0.5\wd\@brx\usebox{\@brx}}}
\newcommand{\rrangle}[1][]{\savebox{\@brx}{\(\m@th{#1\rangle}\)}%
\mathclose{\copy\@brx\kern-0.5\wd\@brx\usebox{\@brx}}}
\makeatother

\begin{document}

\title{Nonreciprocal Charge and Spin Transport Induced by Non-Hermitian Skin Effect in Mesoscopic Heterojunctions}

\author{H. Geng}
\thanks{These authors contributed equally to this work.}
\affiliation{National Laboratory of Solid State Microstructures, School of Physics,
and Collaborative Innovation Center of Advanced Microstructures, Nanjing University, Nanjing 210093, China}

\author{J. Y. Wei}
\thanks{These authors contributed equally to this work.}
\affiliation{National Laboratory of Solid State Microstructures, School of Physics,
and Collaborative Innovation Center of Advanced Microstructures, Nanjing University, Nanjing 210093, China}

\author{M. H. Zou}
\affiliation{National Laboratory of Solid State Microstructures, School of Physics,
and Collaborative Innovation Center of Advanced Microstructures, Nanjing University, Nanjing 210093, China}

\author{L. Sheng}
\affiliation{National Laboratory of Solid State Microstructures, School of Physics,
and Collaborative Innovation Center of Advanced Microstructures, Nanjing University, Nanjing 210093, China}

\author{Wei Chen}
\email{Corresponding author: pchenweis@gmail.com}
\affiliation{National Laboratory of Solid State Microstructures, School of Physics,
and Collaborative Innovation Center of Advanced Microstructures, Nanjing University, Nanjing 210093, China}

\author{D. Y. Xing}
\affiliation{National Laboratory of Solid State Microstructures, School of Physics,
and Collaborative Innovation Center of Advanced Microstructures, Nanjing University, Nanjing 210093, China}


\date{\today }

\begin{abstract}
The pursuit of the non-Hermitian skin effect (NHSE) in various physical systems is of great
research interest. Compared with recent progress in non-electronic systems,
the implementation of the NHSE in condensed matter physics remains elusive.
Here, we show that the NHSE can be engineered in the mesoscopic heterojunctions (system plus reservoir)
in which electrons in two channels of the system moving towards each other have asymmetric coupling
to those of the reservoir. This makes electrons in the system moving forward and in the opposite direction
have unequal lifetimes,  and so gives
rise to a point-gap spectral topology. Accordingly,
the electron eigenstates exhibit NHSE under the open boundary condition,
consistent with the description of the generalized Brillouin zone.
Such a reservoir-engineered NHSE visibly manifests itself as the nonreciprocal
charge current that can be probed by the standard transport measurements.
Further, we generalize the scenario to the spin-resolved NHSE, which can be probed by the
nonreciprocal spin transport. Our work opens a new research avenue
for implementing and detecting the NHSE in electronic mesoscopic systems, which will lead to interesting device
applications.
\end{abstract}

\maketitle

\section{Introduction}
In quantum mechanics,
a closed system is described by a Hermitian
Hamiltonian, which gives rise to
real energy spectrum and unitary evolution
of the system~\cite{Dirac1981}.
In reality, however, physical systems
unavoidably couple to the environment,
which may lead to the exchange of energy,
particles and information~\cite{Breuer2002}.
In many cases, the physics of open systems can still be effectively
described by a non-Hermitian Hamiltonian~\cite{moiseyev2011non,Ashida2020},
which have been widely studied
in various physical systems, such as photonic/optical systems~\cite{miri2019exceptional,Feng2017,Oezdemir2019,Zhu2020},
cold atoms~\cite{Nakagawa2018,Yamamoto2019,Xu2017},
and condensed matter systems~\cite{Nagai2020,Papaj2019,Kozii2017,Shen2018,Bergholtz2019,
San-Jose2016, Cayao2022prb}.
Exotic physical phenomena attributed to non-Hermiticity have been discovered,
such as unidirectional transport~\cite{Abdo2013,Feng2011,Caloz2018,
Abdo2014,Yu2009,Metelmann2015,Fang2017,Peterson2017,Bernier2017,Xu2019,Fleury2014,Sounas2015},
enhanced sensitivity~\cite{Chen2018,Fleury2015,Chen2017,Hodaei2017,Dong2019}, and
single-mode lasing~\cite{Peng2014,Brandstetter2014,Miri2012}, which will lead to
important applications~\cite{Ashida2020}.

Recent progress in non-Hermitian physics is the
discovery of the non-Hermitian skin effect (NHSE)~\cite{Yao2018,Yao2018a,Kunst2018},
in which all the bulk states are driven to the system boundaries under the
open boundary condition (OBC)~\cite{Yao2018,Yao2018a,Kunst2018,Yokomizo2019,Lee2016,Lieu2018,Yin2018,Carlstrm2018,
MartinezAlvarez2018a,Lee2019b,Longhi2019,Longhi2020,Li2020a,Yi2020,Zhang2021,
Esaki2011,Okuma2021,
Okuma2021a,
Zhang2019,Zhou2019,Bergholtz2021,Kawabata2020,Yang2020,
Lee2019a,Borgnia2020,Okuma2020,Zhang2020,
xiao21prl,Xiao2020,Weidemann311,
Li2020,
Helbig2020,Liu2021,Hofmann2020,
Brandenbourger2019,Ghatak2020}.
The NHSE is a unique phenomenon due to the non-Hermiticity, which stems from
the point gap topology of the complex spectrum
under the periodic boundary condition (PBC)~\cite{Lee2019a,Borgnia2020,
Okuma2020, Zhang2020}.
In the presence of the NHSE, the conventional
bulk-boundary correspondence in the topological
band theory fails
and instead, the non-Bloch band theory should be employed~\cite{Yao2018,Yao2018a,Kunst2018,Yokomizo2019}.
Very recently, the NHSE has been observed
in a variety of non-electronic systems, such
as optics~\cite{xiao21prl,Xiao2020,Weidemann311},
acoustics~\cite{zhang21nc,Auregan2017,Christensen2016}, cold atoms~\cite{Liang22prl},
topoelectrical circuit~\cite{Helbig2020,Liu2021,Hofmann2020} and
classical mechanic systems~\cite{Brandenbourger2019,Ghatak2020}.
On the contrary, synthesis and detection of
the NHSE in solid-state systems remain elusive, despite that
the state-of-the-art fabrication techniques
of mesoscopic electronics indicate a plenty of room
for its implementation.

\begin{figure}
\centering
\includegraphics[width=1\columnwidth]{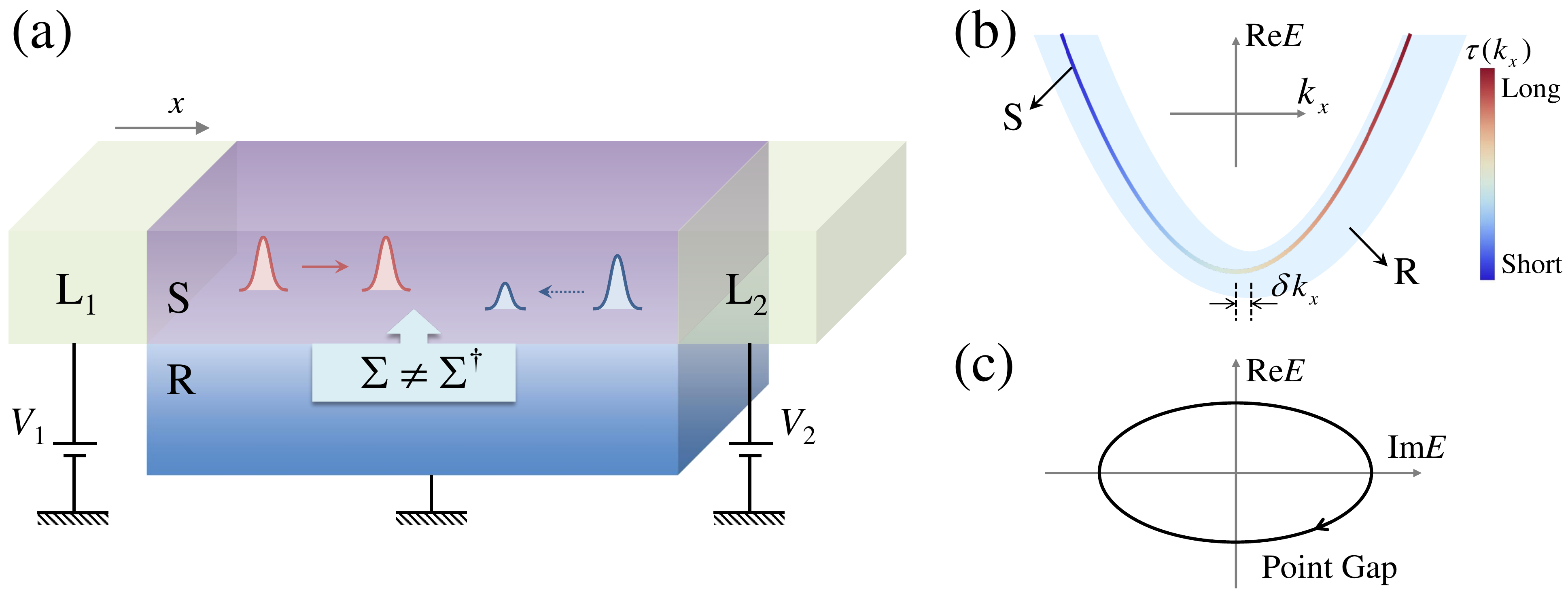}
\caption{Proposed heterojunction composed of the system (S) coupled to
a reservoir (R), the latter introducing a non-Hermitian
self-energy $\Sigma\neq\Sigma^\dag$ to S. Two leads $\text{L}_1$ and $\text{L}_2$ are
coupled only to S and another lead (not shown) is coupled to R. Due to the reservoir-engineered non-Hermiticity,
the propagation of wave packet in S exhibits nonreciprocity in the $x$-direction, \emph{i.e.},
the wave packet propagating leftward decays much faster than that propagating rightward.
(b) Complex band structure in S (colored line) with the electron lifetime
denoted by the color. A relative momentum shift $\delta k_x$
between the band of S and that of R (shadow region)
leads to asymmetric coupling between them.
(c) Complex energy spectrum with the point gap topology.
}\label{fig1}
\end{figure}

In this paper, we propose to engineer and detect the NHSE
in an electronic mesoscopic heterojunction, as shown in Fig.~\ref{fig1}(a),
which is composed of two parts: a system (S)
and a reservoir (R), coupled to each other. Due to the
coherent coupling, S becomes non-Hermitian, and
can be effectively described by the Green's function as
\begin{equation}\label{G}
\begin{split}
\bm{g}^r(\omega)=\frac{1}{\omega-H_{\text{eff}}(\omega)},\ \ H_{\text{eff}}(\omega)=H_{\text{S}}+\Sigma^r_{\text{R}}(\omega),
\end{split}
\end{equation}
where the effective Hamiltonian $H_{\text{eff}}$ of S consists of the
bare Hamiltonian $H_{\text{S}}$ and the retarded self-energy $\Sigma^r_{\text{R}}(\omega)$
due to the coupling between S and R. The self-energy is in general non-Hermitian with $\Sigma^r_{\text{R}}\neq\Sigma_{\text{R}}^{r\dag}$;
see Fig.~\ref{fig1}(a). Given that the dynamics of the electrons in S is governed
by the Green's function $\bm{g}^r$ or
equivalently, the effectively Hamiltonian $H_{\text{eff}}$ in Eq.~\eqref{G},
very interesting non-Hermitian effects can be implemented in S
by properly engineered $\Sigma_{\text{R}}^r$.
Here, we focus on the special type of $\Sigma_{\text{R}}^r$
that can give rise to the NHSE. If S is coupled  to R asymmetrically
for $k_x >0$ and $k_x <0$ in Fig.~\ref{fig1}(b), there will be
unequal lifetimes of electrons in S moving forward and backward.
The resultant $H_{\text{eff}}$
yields a point gap topology in its complex spectrum under the PBC [Fig.~\ref{fig1}(c)],
and accordingly, the wave functions under the OBC
exhibit the NHSE [Figs.~\ref{fig2}(d-f)],
which can be well described by the generalized Brillouin
zone (GBZ) [Fig.~\ref{fig2}(c)].

The great advantage of our proposal is that
the non-Hermitian phenomena can be probed by the
standard transport measurements. To achieve this, two leads L$_{1,2}$
in Fig.~\ref{fig1}(a) are designed to connect \emph{only} to S so that
the current flowing between them provides a direct measure of the
non-Hermitian effects in S. Moreover,
such non-Hermitian physics in S can be
incorporated straightforwardly into the framework
of non-equilibrium Green's function theory for quantum transport~\cite{Datta1995}.
We will show that the point gap topology of the complex spectrum
gives rise to nonreciprocal charge transport between L$_1$ and L$_2$.
Such a non-Hermitian scenario can be generalized to the spin-resolved situation
and lead to nonreciprocal spin transport.

The rest of the paper is organized as follows.
In Sec.~\ref{1d} and Sec.~\ref{1ds}, we show how to engineer both the conventional and
spin-resolved NHSE in the 1D systems by coupling to the
reservoir and discuss the resultant nonreciprocal charge and spin transport phenomena, respectively.
In Sec.~\ref{2d} and Sec.~\ref{2ds}, the main results
of the nonreciprocal charge and spin transport are generalized to 2D systems.
Finally, some discussions and prospects are given in Sec.~\ref{dis}.

\section{Nonreciprocal charge transport in 1D system}\label{1d}
To be concrete, we start with a
1D system arranged in the $x$-direction coupled to a reservoir,
which simulates a nanowire deposited on
a 2D substrate. Assuming the PBC in the $x$-direction,
the whole system can be described by the following Hamiltonian (lattice constant set to unity) as
\begin{equation}\label{1dhameq}
\begin{split}
H &= H_{\text{S}}+ H_{\text{R}} + H_{\text{T}},\\
H_{\text{S}} &= \sum_{k_x} \varepsilon_s(k_x)c_{k_x}^{\dagger} c_{k_x}, \ \ \ H_{\text{T}} = \sum_{k_x} \big( t_0 c_{k_x}^\dagger a_{k_x,0}+ \text{H.c.} \big),\\
H_{\text{R}} &= \sum_{k_x, y=0}^{y=-\infty} \big[\varepsilon_r(k_x) a_{k_x, y}^\dag a_{k_x, y} +(t_r^y a_{k_x, y}^{\dagger} a_{k_x, y-1} + \text{H.c.}) \big].
\end{split}
\end{equation}
Here, $\varepsilon_s(k_x)=2 t_s \cos k_x - U_s$ is the electronic
energy in S measured from its band bottom $U_s$ with $t_s$ the hopping strength,
$\varepsilon_r(k_x)=2 t_r^x \cos \left( k_x +\delta k_x\right)-U_r$
is the $x$-direction energy dispersion in R also measured from its band bottom $U_r$
with $t_r^x$ the relevant hopping strength, and $t_r^y$ is the $y$-direction hopping in R.
$\delta k_x$ describes the
deviation in momentum $k_x$ between the bands of S and R, which mimics the
asymmetric band structures of S and R that generally exist
for different materials, as shown in Fig.~\ref{fig1}(b).
The interface coupling $t_0$ ($\ll t^y_r$) takes place between
S and the outmost layer of R. High quality of the interface
is assumed such that the coupling between
S and R ensures $k_x$ conservation. The Fermi
operators $c_{k_x}$ and $a_{k_x, y}$ correspond to S and R, respectively, the latter being written
in the mixed reciprocal and real spaces for respective directions. The subscript $y=0,-1,\cdots,-\infty$,
which simulates the R connected to another lead in the bottom [cf. Fig~\ref{fig1}(a)].

\begin{figure*}
\centering
\includegraphics[width=2\columnwidth]{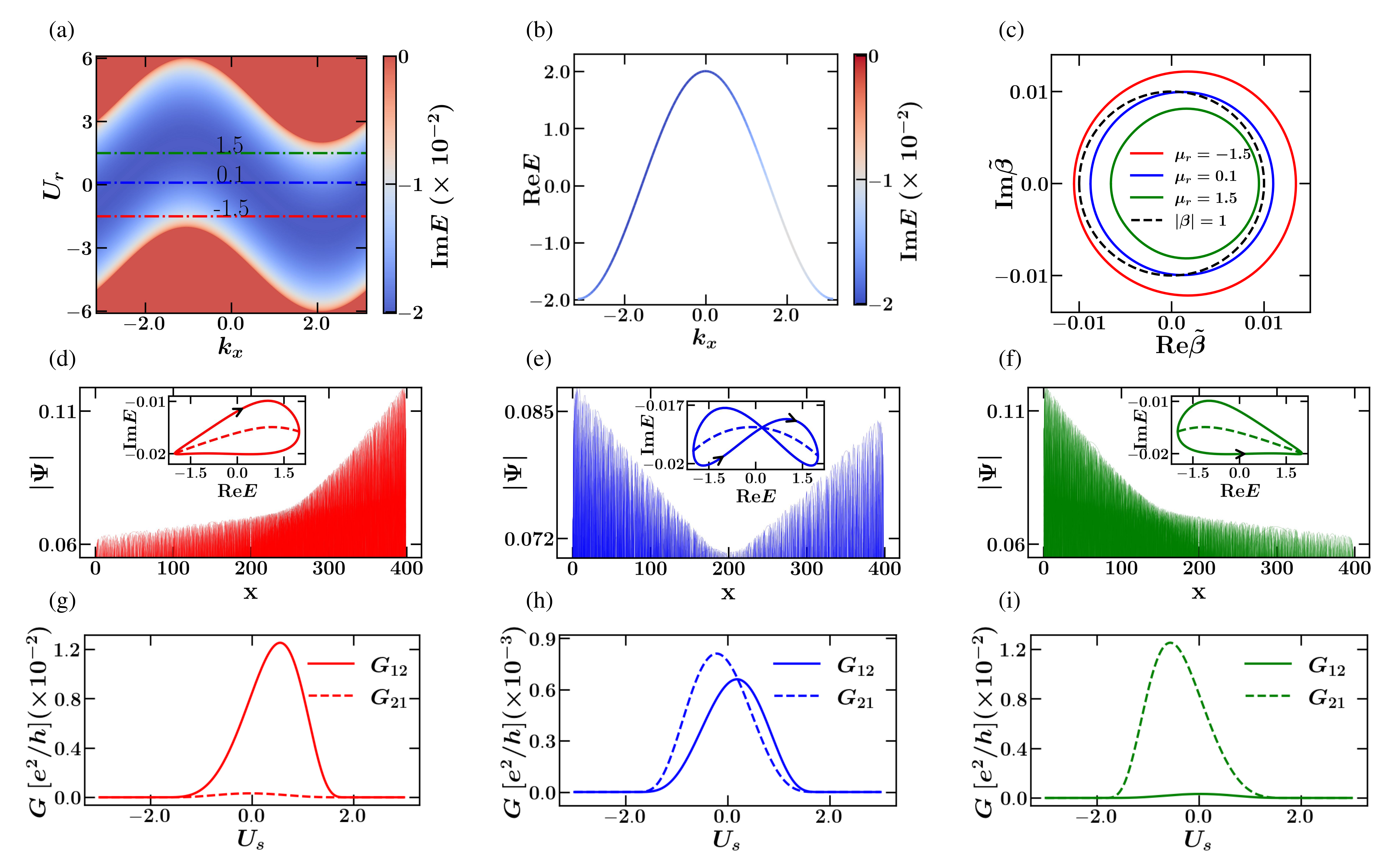}
\caption{(a) $\text{Im}E$ as a function of $U_r$ and $k_x$,
which contains the information of the electron lifetime in S.
(b) Complex energy band with $U_r=1.5$ and $U_s=0$, whose imaginary part is shown by the color.
(c) GBZs calculated by the effective Hamiltonian $H_{\text{eff}}$ of S for $U_r=\pm1.5,0.1$ denoted in (a),
in which $\tilde{\beta}=(|\beta|-0.99)e^{i \text{Arg}[\beta]}$.
(d-f) Normalized eigenfunctions under OBC for various $U_r$ corresponding to those in (c).
Insets: Complex spectral winding under PBC (solid line)
and energy spectrum under OBC (dashed line) with $U_s=0$.
(g-i) Zero-bias differential conductance as a function of $U_s$ that corresponds to
the non-Hermitian properties in (d-f), respectively.
Other parameters are $\omega=0$, $t_s=t_r^x=1$, $t^y_r=2.01$, $t_0=0.1t_r^y$ and $\delta k_x=\pi/3$.
For the results calculated under the OBC, the length of the system in the $x$ direction is $L=400$.
}\label{fig2}
\end{figure*}

By integrating out the reservoir part of Eq.~\eqref{1dhameq},
the effective Hamiltonian $H_{\text{eff}}(\omega)$ in S can be obtained,
with the $k_x$-dependent retarded self-energy as [cf. Appendix~\ref{AppB}]
\begin{equation}\label{srTMEq}
\begin{split}
\Sigma^r_{\text{R}}(k_x,\omega)&=\big[\epsilon-\text{sgn}(\epsilon+1)\sqrt{\epsilon^2-1}\big]t_0^2/t_r^y,\\
\epsilon&= [\omega-\varepsilon_r(k_x)]/(2t_r^y),
\end{split}
\end{equation}
where sgn($\cdot$) is the sign function. Negative imaginary part of the self-energy,
$\text{Im}(\Sigma^r_{\text{R}})<0$, arises as $|\epsilon|<1$
due to the electronic coupling of S and R and accordingly, the
energy $E=\varepsilon_s(k_x)+\Sigma^r_{\text{R}}$ becomes complex. The condition means that for a given $k_x$,
only those electron wave  functions in S with energy $\omega$ satisfying $|\omega-\varepsilon_r|
<2t_r^y$ decay with motion and at the same time appear in R through the interface.
Here, the momentum difference $\delta k_x$ or more generally, the asymmetry in the bands of S and R
is the key ingredient for engineering the
NHSE. It leads to unequal decay of the $\pm k_x$ states in S with
$\text{Im}[E(k_x,\omega)]\neq\text{Im}[E(-k_x,\omega)]$
and so breaks the reciprocity; see Figs.~\ref{fig2}(a) and \ref{fig2}(b). The inverse lifetime of
electrons in S is given by $\tau^{-1}(k_x,\omega)=-\text{Im}[E(k_x,\omega)]$,
which is proportional to velocity $v^R_y(k_x,\omega)$
along the $y$ direction of electrons in R [cf. Appendix~\ref{AppB}].

In what follows we focus on quantum transport at zero bias
($\omega=0$). In Fig.~\ref{fig2}(a) we show $\text{Im}E$
as a function of $k_x$ and $U_r$ with $t_r^y>t_r^x$.
It is found that there are two types of non-Hermitian regions. In region (i) of $|U_r|<2(t_r^y-t_r^x)$,
$E(k_x)$ is entirely complex
for all $k_x$ states and there appears nontrivial winding [insets of
Figs.~\ref{fig2}(d-f)]. If $t_r^y<t_r^x$, there will be no such a full complex-spectrum region.
In region (ii) of $|U_r|\in2(t_r^y-t_r^x,t_r^y+t_r^x)$,
complex energy spectrum appears only for a subset of $k_x$ states,
in which there still be a point gap.
In both non-Hermitian regions, electrons in S moving forward and in the opposite direction
have unequal lifetimes, which leads to the NHSE.
On the contrary, for $|U_r|>2(t_r^y+t_r^x)$, no electron
in S can enter R so that there is no NHSE.

It is known that the ways of spectral winding under the PBC determine
the properties of the skin modes under the OBC~\cite{Okuma2020,Zhang2020}.
To investigate the NHSE, we rewrite Hamiltonian~\eqref{1dhameq} of the whole system
in both directions in real space and calculate the self-energy of S by solving
the surface Green's function of R numerically [cf. Appendix~\ref{APPC}]. The effective
Hamiltonian of S reads $H_{\text{eff}}^{\text{latt}}=\sum_{i,j}t_{ji}c^\dag_jc_i$
with $t_{ji}$ the dressed hopping from site $i$ to $j$. Different from $H_{\text{S}}$,
$H_{\text{eff}}^{\text{latt}}$ involves long-range hopping.
In the non-Hermitian regions (i, ii), we have
$H^{\text{latt}}_{\text{eff}}\neq H^{\text{latt}\dag}_{\text{eff}}$,
since the hopping terms of $t_{ji}\neq t^*_{ij}$ break the reciprocity.
The  eigenfunctions $\Psi$ of $H^{\text{latt}}_{\text{eff}}$ are numerically
solved under the OBC
in the $x$-direction, as shown in Figs.~\ref{fig2}(d-f), exhibiting obvious NHSE.
The full complex spectrum in region (i)
indicate that all eigenstates under the OBC are the skin modes.
The location where these skin modes pile up is determined by
the specific way of the spectral winding, as shown in insets of Figs.~\ref{fig2}(d-f)
with arrows representing the direction in which $k_x$ increases.
The clockwise and
anti-clockwise winding takes place for different $U_r$, and correspond to the skin patterns
stacked on the right and left boundaries, respectively; see
Figs.~\ref{fig2}(d) and \ref{fig2}(f).
There also exists interesting ``$\infty$''-shaped winding~\cite{Okuma2020,Zhang2020},
which gives rise to the skin modes stacked simultaneously on both boundaries, see Fig.~\ref{fig2}(e).
The skin patterns obtained above are consistent
with the description by the GBZ~\cite{Yao2018} in Fig.~\ref{fig2}(c) [cf. Appendix~\ref{APPD}].
In region (ii), the nontrivial winding persists for
the complex energy spectrum [cf. Appendix~\ref{APPE}].

Compared with the energy spectrum, more
information is involved in the complex band structure shown in Fig.~\ref{fig2}(b).
Specifically, its real and imaginary parts will determine the
quantum transport taking place in S.
The asymmetric band structures
result in nonreciprocal transport in S,
which is embodied in the effective Hamiltonian $H_{\text{eff}}$ or the Green's function $\bm{g}^r$.
Such an effective description can be incorporated into the framework of the non-equilibrium Green's function method
to study the transport properties.
The differential conductance between leads L$_{1,2}$ is defined as~\cite{Datta}
\begin{equation}\label{conductanceEq}
\begin{split}
&G_{\alpha\beta}(eV) =\frac{\partial I_\beta}{\partial V_\alpha}= \frac{e^2}{h} \text{Tr} \big[ \bm{\Gamma}_\beta \textbf{G}^{r} \bm{\Gamma}_{\alpha} \textbf{G}^{a}  \big]_{\omega=eV},\\
&\textbf{G}^{r,a}=\bm{g}^{r,a}+\sum_{\alpha=1,2}\bm{g}^{r,a}\Sigma_\alpha^{r,a}\textbf{G}^{r,a},
\bm{\Gamma}_{\alpha}=i(\Sigma_\alpha^r-\Sigma_\alpha^a),
\end{split}
\end{equation}
where subscripts $\alpha,\beta=1,2$  indicate the lead labels.
The full retarded (advanced) Green's function $\textbf{G}^{r}$ ($\textbf{G}^{a}$)
can be solved by the Dyson equation
with self-energy $\Sigma_{\alpha}^{r,a}$ and corresponding linewidth
function $\bm{\Gamma}_{\alpha}$ due to the coupling with lead L$_{\alpha}$.

The NHSE can be detected by the transport signatures between leads L$_{1}$ and L$_{2}$.
The zero-bias differential conductance $G_{\alpha\beta}$ is calculated by Eq.~\eqref{conductanceEq}
on the discrete lattices and its dependence on $U_s$ is plotted in Figs.~\ref{fig2}(g-i).
The one-to-one (up-to-down) correspondence between Figs.~\ref{fig2}(g-i) and  Figs.~\ref{fig2}(d-f)
can be easily understood physically.
The zero-bias conductance for a given $U_s$ reflects
the information at the Fermi level so that
the nonreciprocal conductance varying with $U_s$ can be regarded as a complex spectral tomography.
Although the skin modes are not completely stacked at the boundary,
somewhat different from those in simple non-Hermitian lattices~\cite{Yao2018},
the conductance in Figs.~\ref{fig2}(g,i) exhibits a strong
nonreciprocity with the current flowing in
one direction being much greater than that in the opposite direction.
Such a diode-like effect stems from the unequal lifetimes of electronic states with opposite
momentum, or equivalently, the point gap topology [cf. Fig.~\ref{fig1}].
The degree of nonreciprocity, i.e., the ratio between
the current flowing in the two opposite directions enhances as the length
of the system becomes larger.
The forward direction of the diode is determined by the ways of spectral winding
in Figs.~\ref{fig2}(d,f),
which is also consistent with the direction in which the skin modes are stacked.
For the simple winding, nonreciprocal transport with the same forward direction
takes place for all energies.
On the other hand, the ``$\infty$''-shaped winding in Fig.~\ref{fig2}(e) results in an
energy-dependent nonreciprocity, see Fig.~\ref{fig2}(h).
Specifically, the curves of $G_{12}$ and $G_{21}$ intersect at the energy just
corresponding to the crossing point of the ``$\infty$''-spectrum, as shown in
Figs.~\ref{fig2}(e) and~\ref{fig2}(h). In this case, the nonreciprocal transport effect
is greatly reduced due to the nearly equal $G_{12}$ and $G_{12}$.
The above discussion focuses on the non-Hermitian region (i).
In region (ii), both nonreciprocal and reciprocal transport can be realized
in different energy windows [cf. Appendix \ref{APPE}].

Note that the non-Hermitian physics in S is embodied in
self-energy $\Sigma_{\text{R}}^r$ induced by the coupling to R.
From a mathematical point of view, the calculation of $\Sigma_{\text{R}}^r$
and $\Sigma_{1,2}^r$ induced by  L$_{1,2}$ is on an equal footing.
 Therefore, the conductance
can be calculated in a conventional way~\cite{Datta} by regarding the whole setup (S+R)
in Fig.~\ref{fig1}(a) as a scattering region connecting to
three terminals, which ensures the correctness of our results.
Physically, however, the two types of self-energies play distinctive roles:
proper engineering of $\Sigma_{\text{R}}^r$ yields interesting non-Hermitian effect
while leads L$_{1,2}$ are just used for its detection.

\section{Nonreciprocal spin transport in 1D system}\label{1ds}

\begin{figure*}
\centering
\includegraphics[width=2\columnwidth]{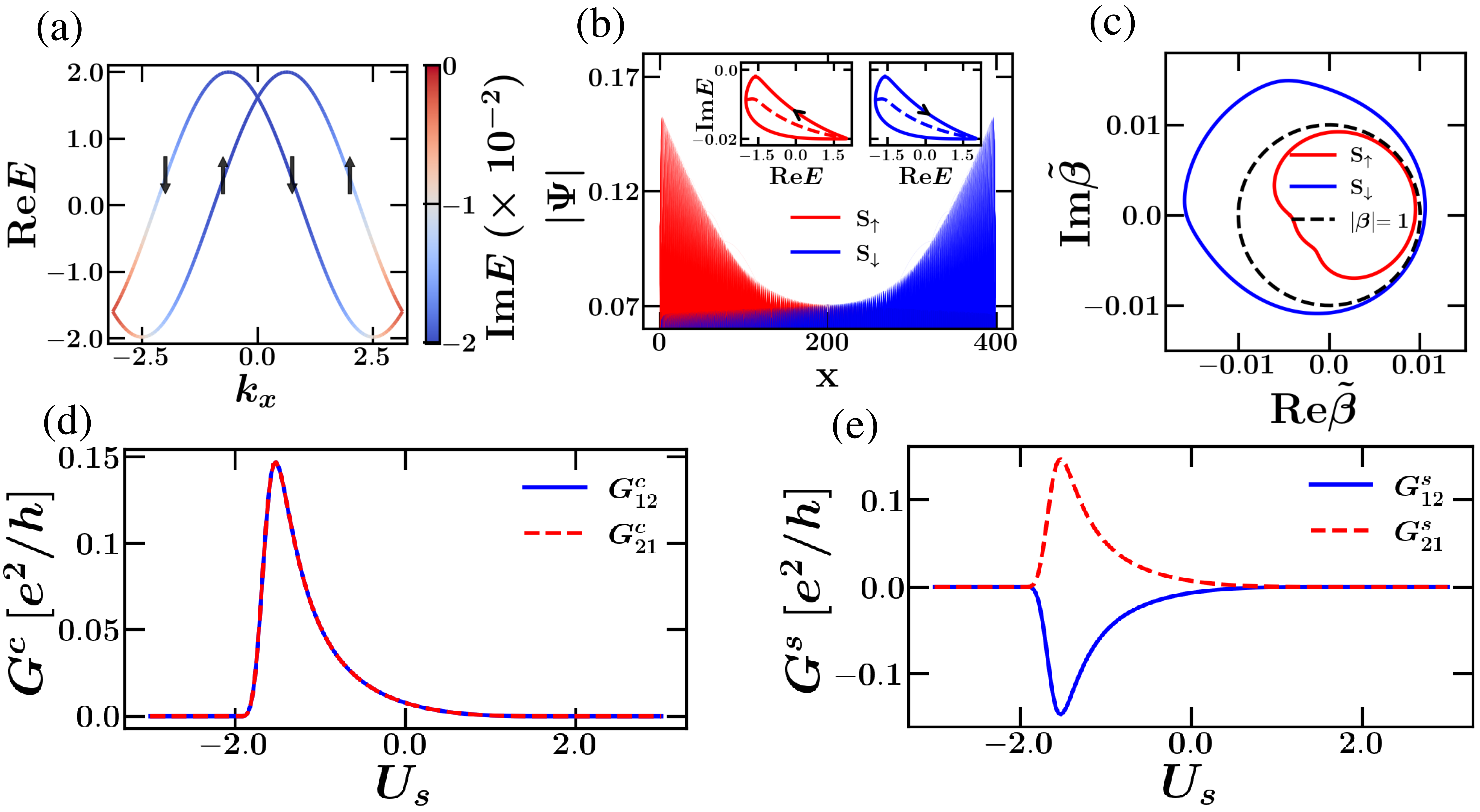}
\caption{(a) Spinful complex energy bands with the arrows denoting
the spin polarizations. (b) Skin modes and spectral winding (insets).
(c) GBZs for both spins.
(d) Charge and (e) spin conductance as a function of $U_s$.
The parameters are $\omega=0$, $\delta k_x=0$, $U_r=1.5$
and $\delta k'_x=\pi/5$ with the others the same
as those in Fig.~\ref{fig2}.
}\label{fig3}
\end{figure*}

The scenario of the NHSE
in heterostructures can be
generalized straightforwardly to the spin-resolved case.
Let's replace $H_\text{S}$ in Eq.~\eqref{1dhameq} by
\begin{equation}\label{1dhamsoceq}
  \tilde{H}_{\text{S}} =
  \sum_{k_x,\sigma=\uparrow,\downarrow}
  \varepsilon_s^{ \sigma}\left(k_x\right)
  c^{\dagger}_{k_x,\sigma}c_{k_x,\sigma}~,
\end{equation}
where $\varepsilon_s^{ \sigma}\left(k_x\right)=
2 t_s \cos \left(k_x+\delta k_x^\sigma\right) - U_s$ is the dispersion
for electrons with spin $\sigma$. The opposite shift of the
momentum $\delta k_x^{\uparrow,\downarrow}=\mp \delta k'_x$ for the two spin states can be
induced by the Rashba spin-orbit coupling. $H_\text{R}$ and $H_\text{T}$ are the same as
those in Eq.~\eqref{1dhameq} except that the spin degeneracy in R is
now considered and $\delta k_x=0$ is taken. Note that
only the relative momentum shift of the bands, rather than their absolute values,
has physical effects.

From the effective Hamiltonian $\tilde{H}_{\text{eff}}$ of S, we numerically obtain
spin-dependent complex band structures, as shown in Figs.~\ref{fig3}(a).
The time-reversal symmetry ensures to have an equal lifetime
for the states with opposite momentum and spin, $i.e.$,
$\tau_\uparrow(k_x,\omega)=\tau_\downarrow(-k_x,\omega)$.
The picture shown in Fig.~\ref{fig3}(a) indicates that
the spin splitting of the bands will lead to spin-resolved nonreciprocity.
Accordingly, the NHSE becomes spin dependent in Fig.~\ref{fig3}(b),
where the spectral winding and the skin-mode accumulation
occur in opposite directions for opposite spin polarizations, which
is also verified by the GBZs in Fig.~\ref{fig3}(c).
The above picture predicts
reciprocal transport for charge but nonreciprocal transport for spin.
The charge conductance $G_{ij}^c= G_{ij}^{\uparrow}+ G_{ij}^{\downarrow}$
and spin conductance $G_{ij}^s= G_{ij}^{\uparrow}- G_{ij}^{\downarrow}$
are plotted in Figs.~\ref{fig3}(d,e). It is found that two different spin states
have equal contribution to $G_{ij}^c$, but opposite contribution to $G_{ij}^s$.
The predicted transport properties are manifested as
$G_{12}^c=G_{21}^c$ and $G_{12}^s=-G_{21}^s$.
Such nonreciprocal spin transport can be used as a spin filter
with the spin polarization being controlled conveniently by the current direction.

\section{Nonreciprocal charge transport in 2D systems and magnetic field effect}\label{2d}

\begin{figure}
\centering
\includegraphics[width=1\columnwidth]{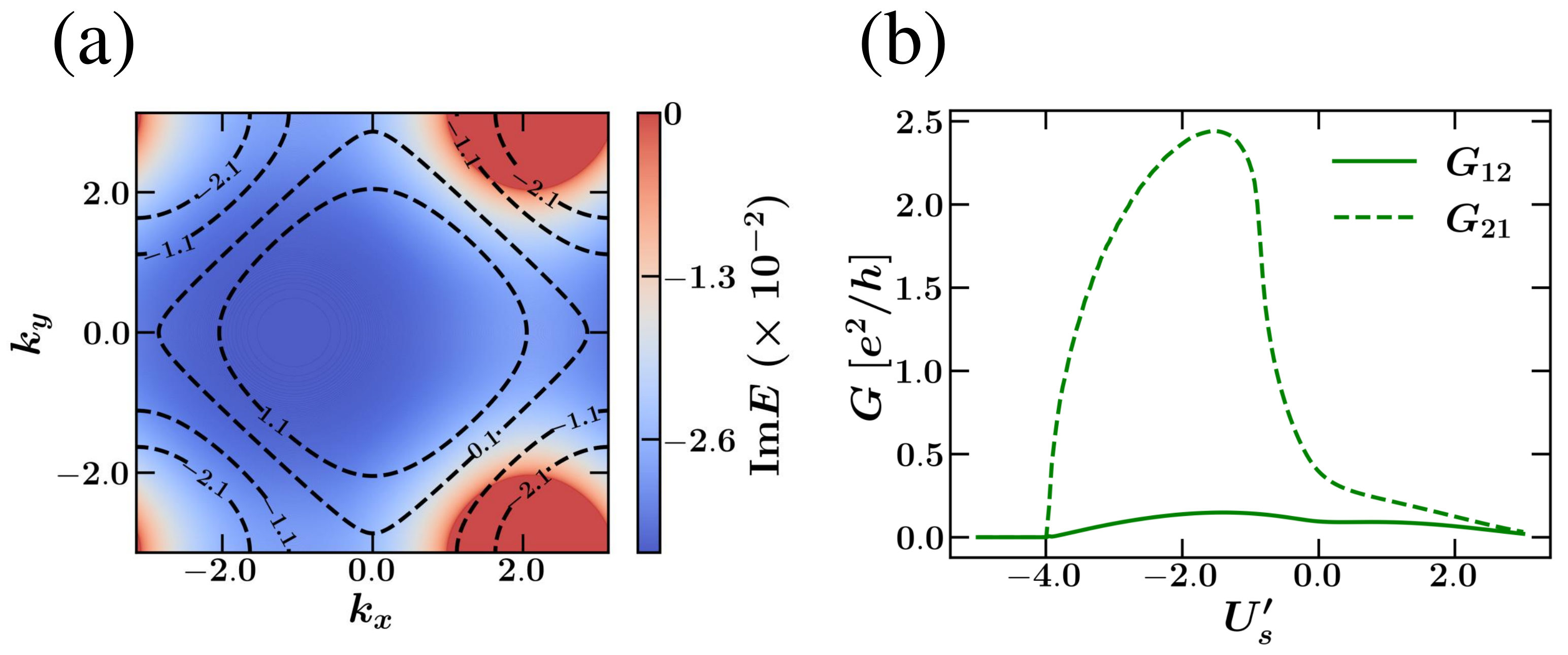}
\caption{(a) $\text{Im}E$ as a function of $k_x$ and $k_y$ with the dashed
lines being the $\text{Re}E$ contours ($E$ is the
eigenvalue of $H_{\text{eff}}^{\text{2D}}$ in Eq.~\eqref{2DHeff}).
(b) Zero-bias differential conductance as a function of $U_s'$.
The relevant parameters are $U_r'=4.1$, $\omega=0$, $t_r'^x=t_r'^y=t_s'=1$, $t_r'^z=3.5$,
$t_0'=0.1t_r'^z$, $B=0$ and $\Delta k_x=\pi/3$. The length ($x$ direction) and width ($y$ direction)
are $L=100$ and $W=200$, respectively.
}\label{2dcharge}
\end{figure}

The general scenario of the
reservoir-engineered NHSE and the resultant nonreciprocal
transport is not restricted to a specific spatial dimension.
In this section, we show that nonreciprocal charge transport can be implemented
in 2D heterostructures as well. We adopt the following lattice Hamiltonian
\begin{widetext}
\begin{equation}\label{2DH}
\begin{split}
H_{\text{S}}^{\text{2D}} &=\sum_{x, y}  \big(t_s'
c_{x+1, y}^\dagger c_{x, y} +
t_s' e^{i B x} c_{x, y+1}^\dagger c_{x, y} -
\frac{U_s'}{2}
c_{x, y}^\dagger c_{x, y} + \text{H.c.}\big),\ \ \ H_{\text{T}}^{\text{2D}} = \sum_{x,y}
\big(t_{0}'  c_{x, y}^\dagger a_{x, y, z=0} +\text{H.c.}\big),\\
    H_{\text{R}}^{\text{2D}} &= \sum_{x,y, z=0}^{z=-\infty}
        \big(t_r'^x  e^{i\Delta k_x} a_{x+1, y, z}^\dagger a_{x, y, z} +
    t_r'^y  e^{i B x} a_{x, y+1, z}^\dagger a_{x, y, z}+t_r'^z a_{x, y, z+1}^\dagger a_{x, y, z} -
    \frac{U_r'}{2} a_{x, y, z}^\dagger a_{x, y, z}+\text{H.c.}\big),\\
\end{split}
\end{equation}
\end{widetext}
where $t_s'$ is the hopping in the system and $t_r'^{x}, t_r'^{y}, t_r'^{z}$ are those in
the reservoir, $U_s'$ and $U_r'$ are the energies
of the band bottoms, $t_0'$ is the interface coupling,
and the momentum deviation $\Delta k_x$ again mimics the asymmetric band structures.
For the later study of the magnetic field effect, we have also introduced
a magnetic field $B$ in the $z$ direction,
which is reflected in the phase factor $e^{i B x}$.

Without a magnetic field, Hamiltonian~\eqref{2DH} has translational invariance in both the
$x$ and $y$ directions. Similar to the 1D case,
we solve the effective Hamiltonian for the system under the PBC in both directions as
\begin{equation}\label{2DHeff}
\begin{split}
&H_{\text{eff}}^{\text{2D}}(\omega, \mathbf{k})= 2 t'_s \cos(k_x)+2t'_s\cos(k_y) -U_s'+\Sigma'^{r}_{\text{R}}(\omega, \mathbf{k})~,\\
&\Sigma'^{r}_{\text{R}}=\Big[ \epsilon'- \text{sgn}\left( \epsilon'+1\right)\sqrt{\epsilon'^2-1}\Big]t_{0}'^{2}/t_r'^z,\\
&\epsilon'= \big[\omega+U'_r-2t_r'^x\cos(k_x+\Delta k_x)-2t_r'^y\cos(k_y)\big]/(2t_r'^z).
\end{split}
\end{equation}
In Fig.~\ref{2dcharge}(a), we plot $\text{Im}E$ as a function of $k_x$ and $k_y$.
One can see that although the Fermi surface of the system is symmetric
about $k_x=0$, $\text{Im}E$ does not. As a result,
nonreciprocal charge transport takes place which is
manifested as $G_{12}\neq G_{21}$ in Fig.~\ref{2dcharge}(b).

\begin{figure}
    \centering
    \includegraphics[width=1\columnwidth]{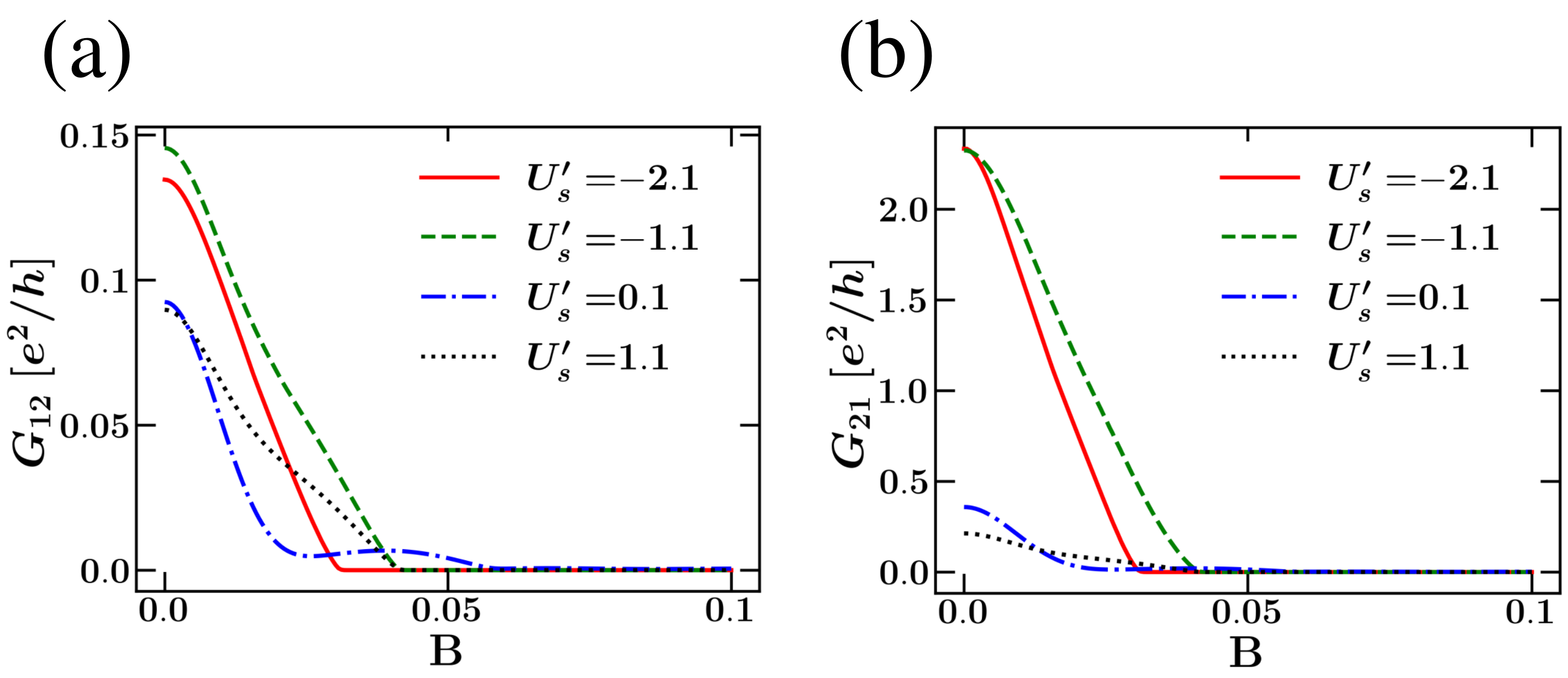}
    \caption{Conductance as a function of the magnetic field $B$.
    The parameters are the same as those in Fig.~\ref{2dcharge}.
    }\label{mag}
\end{figure}

It has been known that a magnetic field will strongly suppress
the non-Hermitian skin effect in 2D systems~\cite{shao22prb,lu21prl}.
In Fig.~\ref{mag}, we plot the conductance as a function of
the magnetic field. One can see that a small magnetic field
strongly suppresses both $G_{12}$ and $G_{21}$
and so the nonreciprocal charge transport disappears.
The sensitivity of the non-Hermitian skin effect
to the magnetic field provides an effective way
for the control of the nonreciprocal charge transport in 2D heterojunctions.

\begin{figure}
    \centering
    \includegraphics[width=1\columnwidth]{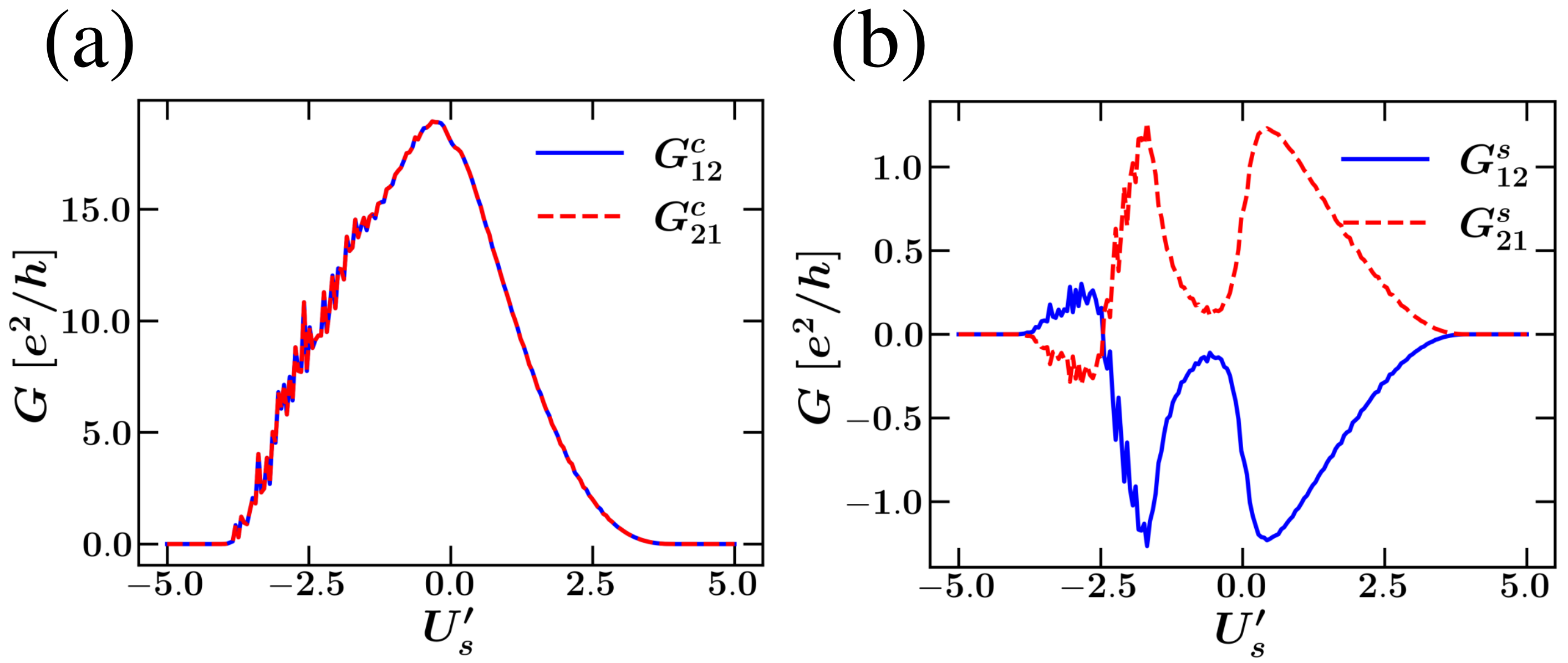}
    \caption{(a) Charge and (b) spin conductance as a function of $U_s'$.
    The parameters are $t_r'^z=3.1, t_0'=0.1t_r'^z, t_{soc}=0.2, L=W=40$
    and the others are the same as those in Fig.~\ref{2dcharge}
    }\label{2dspin}
\end{figure}

\section{Nonreciprocal spin transport in 2D systems}\label{2ds}

In this section, we investigate nonreciprocal spin transport in 2D systems.
Similar to the 1D case, we consider the system to be the 2D electron gas
with Rashba spin-orbit coupling, which is described by the Hamiltonian
\begin{equation}\label{2deg}
\begin{split}
    \tilde{H}_{\text{S}}^{2\text{D}}&=
        \sum_{\bm{k}} c^\dagger_{\bm{k}}h_s(\bm{k})c_{\bm{k}},\\
    h_s(\bm{k})&=\tilde{\varepsilon}_s(\bm{k})+
2t_{soc}( \sin k_x\sigma_y- \sin k_y\sigma_x),
\end{split}
\end{equation}
where $\tilde{\varepsilon}_s(\bm{k})=2t_s(\cos k_x+\cos k_y)-U_s'$
, $t_{soc}$ is the spin dependent hopping due to the Rashba spin-orbit coupling,
the Fermi operator $c_{\bm{k}}=(c_{\bm{k}}^\uparrow, c_{\bm{k}}^\downarrow)^{\text{T}}$
has two spin components and Pauli
matrices $\sigma_{x,y}$ act on the spin. $H_\text{R}^{\text{2D}}$ and $H_\text{T}^{\text{2D}}$ are the same as
those in Eq.~\eqref{2DH} except that the spin degeneracy in the reservoir is
now considered and $\Delta k_x=0$ and $B=0$ are taken.
The self-energy $\Sigma'^{r}_{\text{R}}$ due to the reservoir is the same
as that in Eq.~\eqref{2DHeff} with $\Delta k_x=0$. Here,
$\text{Im} E$ is symmetric about $k_x=0$ but instead, the bare Hamiltonian~\eqref{2deg}
possesses a spin dependent band splitting, which yields
reciprocal charge transport but nonreciprocal spin transport, similar to the 1D case.
Given the specific spin texture of the 2D electron gas,
the nonreciprocity that occurs in the $x$ direction should be most visibly
revealed by the spin current defined with its polarization along the $y$ direction.
The corresponding conductance is denoted by $G_{ij}^{\rightarrow}$ and $G_{ij}^{\leftarrow}$,
whose superscripts $\rightarrow, \leftarrow$ mean the spin components along the $y, -y$ directions,
respectively. In Fig.~\ref{2dspin}, we plot the charge conductance
$G_{ij}^c=G_{ij}^\rightarrow+G_{ij}^\leftarrow$ and spin conductance
$G_{ij}^s=G_{ij}^\rightarrow-G_{ij}^\leftarrow$.
The same as the 1D case, two different spin states
have equal contribution to $G_{ij}^c$, but opposite contribution to $G_{ij}^s$.
The predicted transport properties are again manifested as
$G_{12}^c=G_{21}^c$ and $G_{12}^s=-G_{21}^s$. Interestingly,
the nonreciprocal spin current can have opposite sign
in different energy regions, which stems from the
rich spin texture of the 2D Rashba gas compared with that in the 1D case.

\section{Discussions and prospects}\label{dis}
We discuss the
experimental implementation of our proposal.
The main ingredients, mesoscopic heterostructures with multiple terminals,
are common setups studied in mesoscopic physics~\cite{Datta},
which can be fabricated with mature technology.
To achieve the NHSE, materials with proper band structures and good tunability
by external fields are favorable. For the
spinless NHSE and nonreciprocal charge transport,
the time-reversal symmetry must be broken.
In our example described by Eqs.~\eqref{1dhameq} and \eqref{2DH}, such a
symmetry breaking is introduced by a relative momentum
shift $\delta k_x$ (or $\Delta k_x$) for clarity.
In reality, any band structures of S and R that lead to
unequal lifetimes of electrons counter-propagating in S are sufficient for the NHSE.
For example, the study of
the heterostructures of topological matter has shown that their band structures
have strong external field tunability~\cite{mourik12sci,yuan18prb},
so that both nonreciprocal charge and spin transport is expected to be realized
in these systems. Moreover, the coexistence of the Rashba spin-orbit
coupling and a Zeeman field in the 2D electron systems
can also give rise to an asymmetric coupling to the reservoir
and the resultant NHSE~\cite{tokura2018nonreciprocal,bihlmayer2022rashba}.
For the spin-resolved
NHSE, the time-reversal symmetry does not need to be broken
so that the Rashba spin-orbit coupling is sufficient for
such an effect and the resultant nonreciprocal spin transport. Finally, we remark
that the engineering of the NHSE in mesoscopic systems can
be further extended to other scenarios including
electron-electron, electron-phonon, electron-impurity
scattering and so on~\cite{Nagai2020,Papaj2019,Kozii2017,Shen2018}.
Regardless of different physical origins, the NHSE in the electron systems
can be probed by the transport measurement schemes proposed in this work.

In this paper, we construct the lattice model,
calculate the self-energy and transport properties using the KWANT package~\cite{Groth2014}.

\begin{acknowledgments}
We are very thankful for the helpful discussions with Zhong Wang, Zhesen Yang, Huaiqiang Wang
and Chunhui Zhang. This work was supported by the National Natural Science Foundation of
China under Grant No. 12074172 (W.C.), No.  12222406  (W.C.), No. 11974168 (L.S.) and
No. 12174182 (D.Y.X.), the State Key Program for Basic
Researches of China under Grants No. 2021YFA1400403 (D.Y.X.), the Fundamental Research
Funds for the Central Universities (W.C.), the startup
grant at Nanjing University (W.C.) and the Excellent
Programme at Nanjing University.
\end{acknowledgments}

\appendix
\section{Surface Green's function of the reservoir}\label{A}

To arrive at the retarded self-energy induced
by the semi-infinite reservoir,
we follow the procedure in Ref.~\cite{Wimmer2009}.
The Hamiltonian of the reservoir can be rewritten in the general form of
\begin{equation}\label{eqs1}
    H_{\text{R}}=
    \sum_{y=0}^{-\infty}
    a_{y}^\dagger H_{0} a_{y}+
    a_{y-1}^\dagger H_{-1} a_{y}+a_{y}^\dagger H_{1} a_{y-1},
\end{equation}
where only the $y$ coordinate is shown in the subscript of the Fermi operator $a_{y}$.
Under the open boundary condition (OBC) in the $x$ direction,
$a_y$ is a vector written in real space as $a_y=(a_{x=1, y}, a_{x=2, y},\cdots, a_{x=L, y})^{\text{T}}$;
while under the periodic boundary condition (PBC), the eigenstates in
the $x$ direction can be labeled by $k_x$ and so
$a_y=a_{k_x, y}$. The sites of $y=0$ are the outmost layer of the reservoir
that are coupled to the system.
The matrices $H_0$ and $H_{\pm 1}$ are the unit-cell Hamiltonian and the hopping Hamiltonian
of the reservoir, respectively. Both $H_0$ and $H_{\pm 1}$ are $N_{\text{u.c. }}\times N_{\text{u.c. }}$ square matrices,
with $N_{\text{u.c.}}=L$ ($L$ the length of the system in the $x$ direction) under the OBC and $N_{\text{u.c. }}=1$ under the PBC,
respectively.

We denote the retarded Green's function of the reservoir by $g^r$
and the surface Green's function $\mathcal{G}^r_{\text{R}}=g_{00}^r $
is just the matrix element for the outmost layer.
To obtain $\mathcal{G}^r_{\text{R}}$, we need to solve the quadratic eigenvalue equation
\begin{equation}\label{QuadraticEq}
    \left(\left(\omega-H_{0}\right) \lambda_{n}-H_{1} \lambda_{n}^{2}-H_{-1}\right) \boldsymbol{u}_{n}=0~,
\end{equation}
for a given $\omega$, where $\boldsymbol{u}_{n}$ is the right eigenvector
corresponding to the eigenvalue $\lambda_n$. We also need to calculate the group velocity $v_n$ of the Bloch modes given by
\begin{equation}\label{VelGenEq}
    v_{n} =-\frac{1}{\hbar}
        \text{Im}\left(2 \boldsymbol{u}_{n}^{\dagger}
        H_{1} \lambda_{n}  \boldsymbol{u}_{n}\right)~.
\end{equation}

Solving the quadratic eigenproblem in Eq.~\eqref{QuadraticEq} yields $2N_{\text {u.c.}}$
eigenvalues and eigenvectors, which can be divided into two groups:
\begin{itemize}
    \item $ N_{\text {u.c.}}$ modes moving in the $-y$ direction with
        $ \left|\lambda_{n}\right|<1 $ or $ \left|\lambda_{n}\right|=1 \wedge v_{n}>0 $.
        These eigenvalues are denoted by $ \lambda_{n,<} $
        and the corresponding eigenvectors are
        $ \boldsymbol{u}_{n,<} $ which we collect into the matrix
        $ U_{<}=\left(\boldsymbol{u}_{1,<}, \ldots,
        \boldsymbol{u}_{N_{\text {u.c. }},<}\right) $.
    \item $ N_{\text {u.c.}}$ modes moving in the $y$ direction with
        $ \left|\lambda_{n}\right|>1 $ or $ \left|\lambda_{n}\right|=1 \wedge v_{n}<0 $.
        These eigenvalues are denoted by $ \lambda_{n,>} $
        and the corresponding eigenvectors are
        $ \boldsymbol{u}_{n,>} $ which we collect into the matrix
        $ U_{>}=\left(\boldsymbol{u}_{1,>}, \ldots,
         \boldsymbol{u}_{N_{\text {u.c. }},>}\right) $.
\end{itemize}
Then the surface Green's function can be solved by
\begin{equation}
    \mathcal{G}^r_{\text{R}} H_{-1}=U_{<} \Lambda_{<} U_{<}^{-1} ~,
\end{equation}
and if $H_{-1}$ is invertible we have
\begin{equation}\label{SurfGrEq}
    \mathcal{G}^r_{\text{R}} =U_{<} \Lambda_{<} U_{<}^{-1} H_{-1}^{-1}~,
\end{equation}
where $\Lambda_{<}=\left(
                     \begin{array}{cccc}
                       \lambda_{1,<} &  &  & 0 \\
                        & \lambda_{2,<} &  &  \\
                        &  & \ddots &  \\
                       0 &  &  & \lambda_{N_{\text {u.c. }},<} \\
                     \end{array}
                   \right)
$
is the diagonal matrix composed of the eigenvalues $\lambda_{n,<}$.
With the surface Green's function, the retarded self-energy
can be obtained as
\begin{equation}\label{selfe}
    \Sigma^r_{\text{R}} =
    t_0^2\mathcal{G}^r_{\text{R}}~,
\end{equation}
where $t_0$ is the coupling strength between
the reservoir and the system.

\begin{table*}[htbp]
	\centering  
	\caption{A submatrix of $H^{\text{latt}}_{\text{eff}}$ with the diagonal elements being
    the onsite potential and the upper and lower diagonal elements being the
    nearest-neighbor hopping. The relevant parameters are the same as those in Fig.~\ref{fig2} and $U_r=1.5$.}  
	\label{HeffFig}  
	\begin{tabular}{|c|c|c|c|}
		\hline  
		$7.50\times 10^{-3}-1.69\times 10^{-2}i$ & $1.00-5.51\times 10^{-3}i$ & $-8.03\times 10^{-4}-4.64 \times 10^{-4}i$ & $1.37\times 10^{-4}i$ \\  
		\hline
        $1.00+3.15 \times 10^{-3}i$ & $7.50\times 10^{-3}-1.69\times 10^{-2}i$ & $1.00-5.51\times 10^{-3}i$ & $-8.03\times 10^{-4}-4.64 \times 10^{-4}i$  \\  
		\hline
        $8.03\times 10^{-4}-4.64 \times 10^{-4}i$ & $1.00+3.15 \times 10^{-3}i$ &  $7.50\times 10^{-3}-1.69\times 10^{-2}i$ & $1.00-5.51\times 10^{-3}i$ \\  
		\hline
        $1.37\times 10^{-4}i$ & $8.03\times 10^{-4}-4.64 \times 10^{-4}i$ & $1.00+3.15 \times 10^{-3}i$ & $7.50\times 10^{-3}-1.69\times 10^{-2}i$ \\  
		\hline
	\end{tabular}
\end{table*}

\section{Derivation of Eq.~(3) and the lifetime}\label{AppB}

Under the PBC in the $x$ direction, we have $H_0=2t_x\cos(k_x+\Delta k_x)-U_r$
and $H_1=H_{-1}=t_r^y$ in Eq.~\eqref{eqs1}, making use of the good quantum number $k_x$. The quadratic eigenproblem reduces to
\begin{equation}
\begin{aligned}
    &\lambda^2-2\epsilon\lambda+\epsilon^2=\epsilon^2-1~,\\
    &\epsilon(k_x,\omega)=\frac{\omega+U_r-2t_r^x\cos(k_x+\Delta k_x)}{2t_r^y}~,\\
\end{aligned}
\end{equation}
which yields two roots of $\lambda$:
\begin{equation}
    \lambda_{\pm}=\epsilon\pm \sqrt{\epsilon^2-1}~.
\end{equation}
Without loss of generality, $t_r^{x,y}$ are chosen to be real and so is $\epsilon$.
If $|\epsilon|\le 1$,
the roots contain an imaginary part
so that $|\lambda_\pm|=1$,
which corresponds to the propagating modes in the reservoir.
To get the retarded surface Green's function, we need to pick up
those modes propagating in the $-y$ direction by their group velocity.
Here, the group velocity in Eq.~\eqref{VelGenEq} reduces to
\begin{equation}\label{Eqvel}
    v_{\pm} = - \frac{1}{\hbar}\text{Im}\left( 2 t_r^y \lambda_{\pm}\right)~,
\end{equation}
and for $t_r^y>0$, $\lambda_{-}$ corresponds to the outgoing modes that we want.
Then the retarded surface Green's function of the reservoir is
\begin{equation}\label{G1d}
    \mathcal{G}^r_{{\text{R}}} =
    \lambda_{-} /t_r^y~.
\end{equation}
If $|\epsilon|> 1$, the roots are real numbers, and only those evanescent modes with $|\lambda|<1$
are relevant. In this case, the surface Green's function can be expressed as
\begin{equation}
    \mathcal{G}^r_{{\text{R}}}=
    \lambda_{\text{sgn}(-\epsilon)}/t_r^y~.
\end{equation}
Inserting $\mathcal{G}^r_{{\text{R}}}$ into Eq.~\eqref{selfe} yields the self-energy
\begin{equation}\label{sr1dexm}
    \Sigma^r_{\text{R}}(k_x,\omega) =
    \begin{cases}
        \left( \epsilon- \text{sign}\left( \epsilon\right)
        \sqrt{\epsilon^2-1} \right)t_0^2/t_r^y& |\epsilon| \geq 1~,\\
        \left(\epsilon -i\sqrt{1-\epsilon^2} \right)t_0^2/t_r^y& |\epsilon|< 1~,
    \end{cases}
\end{equation}
which can be incorporated into the unified form of Eq.~(3).

From Eqs.~\eqref{selfe}~\eqref{Eqvel} and~\eqref{G1d}, one
can obtain the relation between the lifetime of the quasiparticle in the system and the velocity in the reservoir as
\begin{equation}
    \frac{1}{\tau(k_x,\omega)}= -\text{Im}\Big[\frac{\Sigma^r_{\text{R}}(k_x,\omega)}{\hbar}\Big]=
        \frac{1}{2}\left(\frac{t_0}{ {t_r^y}}\right)^2 v_{y}^{R}(k_x,\omega)~,
\end{equation}
where $v_{y}^{\text{R}}=v_{-}$.
One can see that the lifetime of the quasiparticle in the system
is inversely proportional to the electron velocity along the $y$
direction in the reservoir.

\section{Retarded self-energy in real space and nonreciprocal hopping}\label{APPC}

\begin{figure*}
\centering
\includegraphics[width=2\columnwidth]{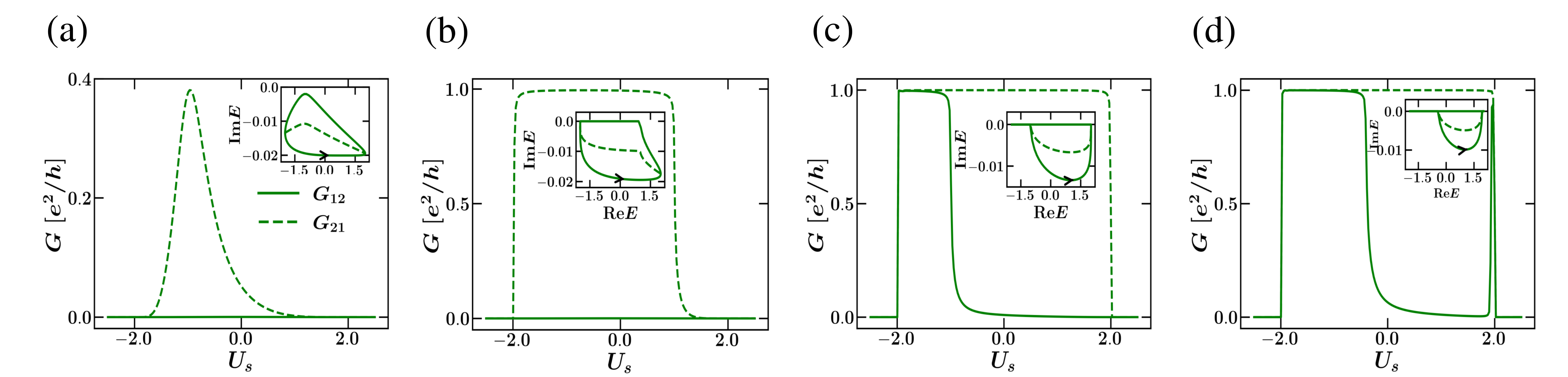}
\caption{ Comparison between the transport properties
and complex spectrum [insets] in (a) non-Hermitian region (i)
and (b-d) non-Hermitian region (ii). $U_r=2.0, 3.0, 5.0, 5.5$ in (a-d), respectively.
The other parameters are the same as those in Fig.~\ref{fig2}.
}\label{1DTranFig}
\end{figure*}

Under the OBC, the matrices $H_0$, $H_{\pm 1}$ in
Eq.~\eqref{eqs1} are
\begin{equation}\label{H0Eq}
\begin{split}
    H_0&=\left(\begin{array}{cccc}
        -U_r & t_r^x& 0&\cdots\\
        t_r^x& -U_r&t_r^x&\cdots\\
        \vdots&\vdots&\vdots&\vdots\\
        \cdots&0&t_r^x&-U_r
    \end{array}\right)_{{L}\times{L}},\\
    H_{\pm 1}&=\left(\begin{array}{cccc}
        t_r^y & 0& 0&\cdots\\
        0 & t_r^y &0&\cdots\\
        \vdots&\vdots&\vdots&\vdots\\
        \cdots&0&0&t_r^y
    \end{array}\right)_{{L}\times{L}},
    \end{split}
\end{equation}
with $L$ the length of the lattice in the $x$ direction.
Following the procedure illustrated in Sec.~\ref{A}, we obtain the
retarded self-energy matrix $(\Sigma_{\text{R}}^r)_{{L}\times{L}}$ under the OBC.
Combining the bare lattice Hamiltonian of the system,
\begin{equation}\label{HsEq}
    H_{\text{S}}^{\text{latt}}=\left(\begin{array}{cccc}
        -U_s & t_s& 0&\cdots\\
        t_s& -U_s&t_s&\cdots\\
        \vdots&\vdots&\vdots&\vdots\\
        \cdots&0&t_s&-U_s
    \end{array}\right)_{{L}\times{L}}~,
\end{equation}
and the self-energy $(\Sigma_{\text{R}}^r)_{{L}\times{L}}$ yields the effective non-Hermitian lattice Hamiltonian
$H_{\text{eff}}^{\text{latt}}=H_{\text{S}}^{\text{latt}}+(\Sigma_{\text{R}}^r)_{{L}\times{L}}$, which can be expressed in the general form
$H_{\text{eff}}^{\text{latt}}=\sum_{i,j}t_{ji}c^\dag_jc_i$. In the non-Hermitian regions (i, ii)
discussed in Sec.~\ref{1d}, we have
$H^{\text{latt}}_{\text{eff}}\neq H^{\text{latt}\dag}_{\text{eff}}$.
For clarity, we exemplify the matrix elements of $H_{\text{eff}}^{\text{latt}}$
in Table ~\ref{HeffFig}, where the effective Hamiltonian
contains long-range hopping and importantly, the hopping terms satisfying
$t_{ji}\neq t^*_{ij}$ break the reciprocity and lead to the non-Hermitian skin effect.

\section{Calculation of the generalized Brillouin zone}\label{APPD}
The generalized Brillouin zone (GBZ)
is the key concept of non-Bloch band theory, which provides an
effective description of the non-Hermitian skin effect.
Here, we illustrate the numerical procedure for the calculation of the GBZ following Refs.~\cite{Yao2018,Yang2020}:
\begin{itemize}
\item Rewrite the eigenvalue equation
    $H_{\text{eff}}\left(\omega,k_x\right)-E=0$ into
    $H_{\text{eff}}\left(\omega,\beta\right)-E=0$ with the parameter defined by
    $\beta=e^{ik_x}$.
\item Solve the eigenvalues $E_{\text{OBC}}$s of the lattice Hamiltonian $H_{\text{eff}}^{\text{latt}}$ under the OBC
    with $\omega=0$.
\item Insert $E_{\text{OBC}}$ into the equation
    $H_{\text{eff}}\left(\omega,\beta\right)-E_{\text{OBC}}=0$ and find
    two roots $\beta_{1,2}$ of $\beta$ for each $E_{\text{OBC}}$. Plot all
    the roots on the complex plane, which
    give the GBZ in Fig.~\ref{fig2}(c).
\end{itemize}

\section{Results for non-Hermitian region (ii)}\label{APPE}

For $t_r^y>t_r^x>0$, the system is in
non-Hermitian region (ii) if $|U_r| \in 2 (t_r^y-t_r^x,t_r^y+t_r^x)$,
and for $t_r^y<t_r^x$, only non-Hermitian region (ii) exists.
The main difference between the two non-Hermitian regions
is that the energy spectrum in region (i) is entirely complex
while that in region (ii) also contains real parts, which
can be seen in Fig.~\ref{fig2}(a).
This fact is reflected by the spectral loop
under the PBC, in which certain segments of the loop lie
on the real axis; see Figs.~\ref{1DTranFig}(b-d). In those
states with real energies, electrons can propagate
without loss and give rise to quantized conductance; see $G_{21}$
in Figs.~\ref{1DTranFig}(b-d). Meanwhile, whether nonreciprocal
transport takes place or not depends on the energy, which can
still be inferred from the complex spectrum. Each energy
$\text{Re}E$ has two states [$\pm k_x$ in Fig.~\ref{fig2}(b)],
which may split into two branches due to unequal $\text{Im}E$ and give rise to
a point gap or may coincide at the real axis with $\text{Im}E=0$ for both states.
For the former case, the conductance satisfies $G_{12}\neq G_{21}$
and exhibits strong nonreciprocity while for the latter case, one has $G_{12}=G_{21}$
and the nonreciprocity disappears; see the results in Figs.~\ref{1DTranFig}(c,d) in different
energy regions.

\bibliographystyle{apsrev4-1}

%

\end{document}